\documentclass[twocolumn]{book}
\usepackage[dvips]{graphicx,color}
\usepackage{makeidx,universe}
\usepackage{amsmath,amssymb}

\makeauthorindex

\BookTitle{Proceedings of the XXIX PHYSICS IN COLLISION}

\CopyRight{\copyright 2009 by Universal Academy Press, Inc.}

\begin{document} 

\pagenumbering{arabic}

\chapter{%
{\LARGE \sf
Lattice QCD in Collision} \\
{\normalsize \bf 
  Shoji Hashimoto$^{1,2}$ 
} \\
{\small \it \vspace{-.5\baselineskip}
(1) KEK Theory Center, Institute of Particle and Nuclear Studies,
     High Energy Accelerator Research Organization (KEK), 
      Tsukuba 305-0801, Japan\\
(2) School of High Energy Accelerator Science, 
     the Graduate University for Advanced Studies (Sokendai), 
      Tsukuba 305-0801, Japan 
}
}


\AuthorContents{S.\ Hashimoto}

\AuthorIndex{Hashimoto}{S.}

\baselineskip=10pt 
\parindent=10pt    

\section*{Abstract} 

In this talk, I present my personal view on the status of lattice QCD
calculations. 
I emphasize the role played by the chiral perturbation theory
($\chi$PT) in analyzing the lattice data of various physical
quantities, including chiral condensate, topological susceptibility,
and pion mass and decay constant. 
I then discuss on the status of determination of fundamental
parameters and other quantities of phenomenological interest.

\vspace*{4mm}
\section{Introduction} 
It was already more than 30 years ago that Quantum Chromodynamics
(QCD) was proposed as the fundamental theory of strong interaction.
Since then, there have been an enormous number of experimental data
that support QCD including its quantum effects.
Those experiments are essentially measuring perturbative aspects of
QCD at high energy, while the non-perturbative dynamics at low energy
still remains as a difficult problem.
Solving QCD is difficult mainly because its vacuum is so complicated.
There is no single dominant gauge field configuration; it is not
completely random either.
Since hadrons are floating on this vacuum,
an understanding of the QCD vacuum is a prerequisite 
to calculate hadron properties starting from first principles.

Lattice QCD is one of the methods to regularize ultraviolet
divergences in QCD.
Unlike the commonly used dimensional regularization, whose definition
involves perturbation theory, lattice QCD is valid in both ultraviolet
and infrared regimes. 
As the regulated theory is mathematically well-defined, direct
numerical calculation of the path integral that defines QCD is possible.
Thus, lattice QCD provides a powerful method for first-principles
calculation of QCD including its non-perturbative dynamics at low energy.
It requires huge computational resources, especially to incorporate
the fermion loop effects in the path integral.
Since 1980s, lattice QCD has been using the high-end supercomputers
available at the time.
It is remarkable that researchers of lattice QCD have even developed
machines that lead the entire supercomputing. 

The state-of-the-art of lattice QCD simulations could be summarized as
follows: 
{\it
``Realistic simulation of QCD to study the static properties of
  hadrons is now feasible''.
}
It means that the inclusion of up, down and strange sea quarks has
become a standard at small enough lattice spacing ($a\lesssim$ 0.1~fm)
on large enough volume ($L\gtrsim$ 2.5~fm) to hold a single hadron.
The low-energy hadron spectrum, for instance, has been well reproduced
by several lattice groups.

Although the recent progress of lattice QCD is so impressive, it still
has many limitations.
One of those is the multi-scale problems.
The scales that may enter in the QCD phenomena are widely ranged:
up and down quark masses ($\sim$ 5~MeV), strange quark mass ($\sim$
100~MeV), the QCD scale ($\sim$ 300~MeV), charm quark mass ($\sim$
1.5~GeV), and bottom quark mass ($\sim$ 4.5~GeV).
Treating light quarks requires more computational costs that grows as
$1/m_q^{2-3}$ with $m_q$ the light quark mass; 
reducing the lattice spacing to treat heavy quarks
needs more resources that typically scales as $1/a^7$ with $a$ the
lattice spacing.
Therefore, the best strategy for practical applications is to use
``effective theories'' such as the chiral perturbation theory
($\chi$PT) for light quarks and the heavy quark effective theory
(HQET) for heavy quarks.
The light and heavy quark masses for which these effective theories are
valid have to be carefully investigated using lattice QCD calculations.

This talk is not a comprehensive review of the field.
Rather, I would like to present my personal view of the status of
lattice QCD.
I start the discussion from a study of the fundamental property of the
QCD vacuum in Section~2.
Then I summarize the calculations of some interesting phenomenological
quantities of light hadrons with special emphasis on the convergence
property of the chiral expansion
(Section~3).
Determination of the fundamental parameters of QCD, such as the strong
coupling constant and quark masses is of particular importance, that I
discuss in Section~4.
My discussion on heavy flavor physics is brief
(Section~5).
My summary and perspective are in the last section.

\vspace*{4mm}
\section{Chiral symmetry breaking}
\label{sec:Chiral_symmetry_breaking}

\subsection{Chiral symmetry and lattice QCD}
One of the fundamental properties on the QCD vacuum is the spontaneous
breaking of chiral symmetry.
Even before QCD, many important properties of low-energy hadrons, such
as the GMOR relation and other soft pion theorems, were discovered
based on the PCAC relation and current algebra. 
From the modern perspective, they are derived from the chiral
effective theory, which is constructed assuming the spontaneous chiral
symmetry breaking.
For the thorough understanding of strong interaction, it is therefore
a crucial step to establish a link between QCD and chiral effective
theory. 

Chiral symmetry of course plays a key role in the understanding
of chiral symmetry breaking.
In the flavor non-singlet sector of chiral symmetry, pions arise as
the Nambu-Goldstone boson associated with the spontaneous symmetry
breaking, while in the flavor-singlet sector the chiral symmetry is
violated by the axial anomaly and is related to the topology of
non-Abelian gauge theory.
There are near-zero modes of quarks associated with the topological
excitations; their accumulation in the vacuum leads to the
symmetry breaking in the flavor non-singlet sector as indicated by the
Banks-Casher relation \cite{Banks:1979yr}. 
Therefore, the initial setup to study the chiral symmetry breaking
should preserve the flavor singlet and non-singlet chiral symmetries.

There is a problem in realizing the chiral symmetry on the lattice.
The conventional Wilson-type fermions violate the chiral symmetry at
the action level, and 
the discrimination between the physical effect of symmetry breaking
and the lattice artifact is not clear.
On the other hand, the staggered fermions have a chiral symmetry but
break the flavor symmetry.
With these lattice fermions, the continuum limit has to be taken
before analyzing the data using the continuum chiral effective theory. 

The domain-wall \cite{Kaplan:1992bt,Shamir:1993zy,Furman:1994ky} and
overlap \cite{Neuberger:1997fp,Neuberger:1998wv} fermions solve this
problem in a theoretically clean manner.
Since they satisfy a modified version of chiral symmetry at finite
lattice spacing \cite{Luscher:1998pq}, the continuum-like
axial-Ward-Takahashi identities are hold on the lattice.
(The chiral symmetry is exact for the overlap fermion; for the
domain-wall fermion the chiral symmetry is restored in the limit of
infinite size of the fifth dimension, but small violation remains in
practical simulations.)
This means that the soft-pion theorems are all satisfied on the
lattice just as in the continuum theory, as far as the chiral symmetry
is spontaneously broken.
The use of $\chi$PT is therefore justified at any finite lattice
spacing $a$. 
(For the Wilson and staggered fermions, one has to include the terms
that represent the violation of chiral symmetry.)
Although the numerical cost is substantially high compared to the
Wilson or staggered fermions, these fermion formulation should
therefore be used when the chiral symmetry is crucial.

The spontaneous chiral symmetry breaking is probed by the chiral
condensate $\langle\bar{q}q\rangle$.
Its lattice calculation is challenging because the scalar density
operator $\bar{q}q$ has a power divergence of the form $m_q/a^2$ as
the cutoff $1/a$ goes to infinity.
The massless limit has to be taken to obtain physical result.
(When the chiral symmetry is violated from the outset as in the
Wilson-type fermions, the divergence is even stronger $\sim 1/a^3$.)
On the other hand, the condensate vanishes in the massless limit,
when the space-time volume is kept finite.
Therefore, the proper order of the limits is to take the infinite
volume limit first and then the massless limit, which is called the
thermodynamical limit. 

\subsection{Spectral density and chiral condensate}
The problem of the ultraviolet divergence can be avoided by focusing
on low-lying eigenmode spectrum of the Dirac operator.
As indicated by the Banks-Casher relation \cite{Banks:1979yr},
the chiral symmetry breaking is induced by an accumulation of
low-lying eigenstates of quark-antiquark pair
\begin{equation}
  \label{eq:Banks-Casher}
  \lim_{m\to 0}\lim_{V\to\infty}\rho(\lambda=0) = \frac{\Sigma}{\pi},
\end{equation}
where $\rho(\lambda)$ denotes the eigenvalue density of the Dirac
operator, 
$\rho(\lambda)\equiv
(1/V)\sum_k\langle\delta(\lambda-\lambda_k)\rangle$.
The expectation value $\langle\cdots\rangle$ represents an ensemble
average and $k$ labels the eigenvalues of the Dirac operator on a
given gauge field background.
On the right hand side of (\ref{eq:Banks-Casher}), 
$\Sigma$ is the chiral condensate, $\Sigma=-\langle\bar{q}q\rangle$,
evaluated in the massless quark limit.
In the free theory, we expect a scaling $\rho(\lambda)\sim \lambda^3$
for a dimensional reason and thus $\rho(0)=0$.
The relation (\ref{eq:Banks-Casher}) implies that the spontaneous
chiral symmetry breaking characterized by non-zero $\Sigma$ is
related to the number of near-zero modes in a given volume.

\begin{figure}[t]
  \centering
  \includegraphics[width=.46\textwidth]{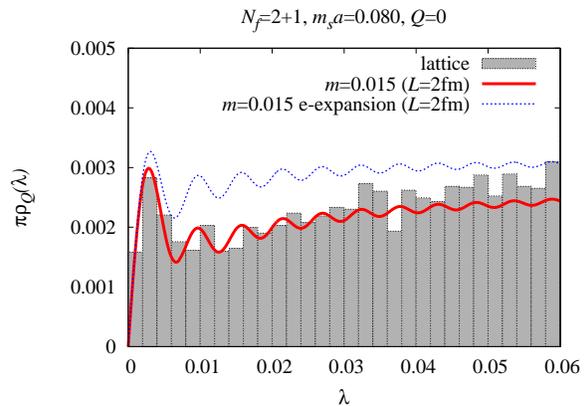}
  \caption{
    Spectral density of the Dirac operator calculated by the JLQCD
    collaboration (histogram).
    The curve represents a next-to-leading order $\chi$PT calculation
    \cite{Damgaard:2008zs}. 
    The pion mass is around 300~MeV and the system in the $p$-regime.
  }
  \label{fig:spectral_density}
\end{figure}

Based on $\chi$PT, more detailed forms of $\rho(\lambda)$ at finite
$\lambda$, $V$ and $m$ have been obtained.
This provides a theoretical basis to control the scaling under the
change of these parameters.
A recent simulation of the JLQCD collaboration gave the spectral
density with 2+1 flavors of dynamical overlap fermions
\cite{Fukaya:2009fh}.
(An earlier analysis is in \cite{Fukaya:2007fb,Fukaya:2007yv}).
One of the results is shown in Figure~\ref{fig:spectral_density},
which is obtained at lattice spacing $\sim$ 0.11~fm and lattice volume
$\sim$ (1.8~fm)$^3\times$(5.4~fm).
The plot compares the result with a $\chi$PT calculation at the
next-to-leading order \cite{Damgaard:2008zs}.
The shape of the spectrum is nicely reproduced, and the height
determines the chiral condensate at the given quark mass.

A similar analysis is attempted using the Wilson fermion in
\cite{Giusti:2008vb}.
Due to the explicit violation of chiral symmetry, the spectrum of the
near-zero modes is distorted.
Relatively larger eigenvalues are, therefore, mainly used in the
comparison with $\chi$PT.

\begin{figure}[t]
  \centering
  \includegraphics[width=.46\textwidth]{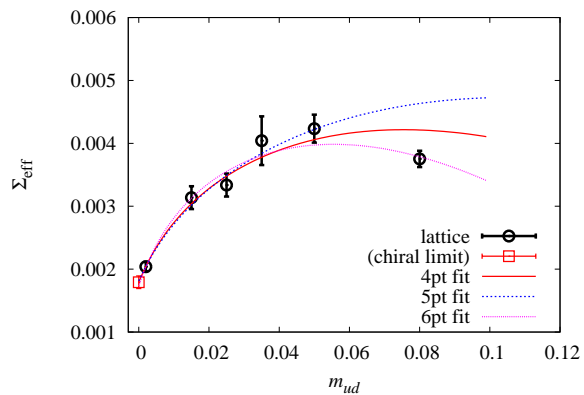}
  \caption{
    Chiral condensate as a function of sea quark mass. 
    Results from the JLQCD collaboration.
  }
  \label{fig:chiral_condensate}
\end{figure}

An extrapolation of the chiral condensate $\Sigma(m_{ud},m_s)$ to the
chiral limit of up and down quark masses is shown in
Figure~\ref{fig:chiral_condensate}.  
The lattice data of the JLQCD collaboration show a curvature, which is
essentially a pion-loop effect as predicted by $\chi$PT \cite{Gasser:1983yg}
\begin{eqnarray}
  \label{eq:chiral_condensate}
  \lefteqn{
    \Sigma(m_{ud},m_s) = \Sigma(0,m_s) \times
  }
  \nonumber\\
  &&
  \left[
    1-\frac{3M_\pi^2}{32\pi^2F^2}\ln\frac{M_\pi^2}{\mu^2}
    +\frac{32 L_6 M_\pi^2}{F^2}
  \right].
\end{eqnarray}
The data point close to the chiral limit, which is in the so-called
$\epsilon$-regime, is helpful to identify this curvature.
A preliminary result in the chiral limit of up and down quarks is 
$\Sigma^{\overline{\mathrm{MS}}}(0,m_s;\mathrm{2~GeV}) =
[243(4)(^{+16}_{-\ 0})\mathrm{~MeV}]^3$.

\subsection{Topological susceptibility}

\begin{figure}[t]
  \centering
  \includegraphics[width=.48\textwidth,clip=on]{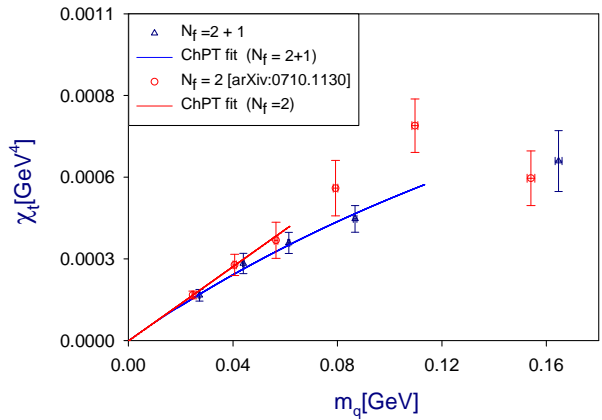}
  \caption{
    Topological susceptibility $\chi_t$ as a function of sea quark
    mass. 
    Lattice data are from \cite{Aoki:2007pw,Chiu:2008kt}.
  }
  \label{fig:chit}
\end{figure}

The topology of gauge field configuration plays a key role in the
accumulation of the near-zero modes. 
The amount of topological excitations is measured by the topological
susceptibility $\chi_t=\langle Q^2\rangle/V$, where $Q$ is the global
topological charge in a given volume $V$.
An expectation from $\chi$PT is that $\chi_t$ behaves as
$\chi_t=m\Sigma/N_f$ with $m$ the quark mass of $N_f$ degenerate
flavors \cite{Leutwyler:1992yt}.
(There is also a formula for non-degenerate quark masses.)
A lattice result by the JLQCD and TWQCD collaboration
\cite{Aoki:2007pw,Chiu:2008kt} obtained with the overlap fermion is
shown in Figure~\ref{fig:chit}.
The results from two-flavor and 2+1-flavor QCD are well described by
$\chi$PT.
This precise calculation was made possible by a new method to
calculate $\chi_t$ through a topological charge density correlation 
on gauge configurations at fixed global topological charge
\cite{Aoki:2007ka}. 

Through these studies, the spontaneous breaking of chiral symmetry is
well established using the first-principles calculation of lattice QCD.
The exact chiral symmetry provided by the overlap fermion played a
crucial role there.
The $\chi$PT is confirmed to be valid near the chiral limit for
fundamental quantities such as the chiral condensate and topological 
susceptibility. 
The next question would be how much the region of $\chi$PT is extended 
towards wider applications and larger values of pion masses/momenta.

\vspace*{4mm}
\section{Light hadron phenomenology}
\label{sec:Light_hadron_phenomenology}

\subsection{Convergence of chiral expansion}
At Lattice 2002, the annual conference on lattice field theory, there
was a panel discussion on the issue of chiral extrapolation of lattice
data \cite{Bernard:2002yk}.
The problem at that time was that the curvature due to the chiral
logarithm of the form $m_\pi^2\ln m_\pi^2$ was not visible in lattice
data for any physical quantity.
This was mainly because the quark mass in the dynamical fermion
simulations at that time was too large that pion mass was above
500~MeV, which is presumably out of the range of $\chi$PT.
This lead to a large systematic error in the chiral extrapolation.

Since then, by the development of algorithms and machines, the pion
mass in the lattice simulations has been reduced down to 200--300~MeV.
For a summary of recent large-scale simulations, see the plenary talk
by Scholz at Lattice 2009 \cite{Scholz:2009yz}.
Now, it is therefore the time to investigate the convergence property
of the chiral expansion.

\begin{figure}[tb]
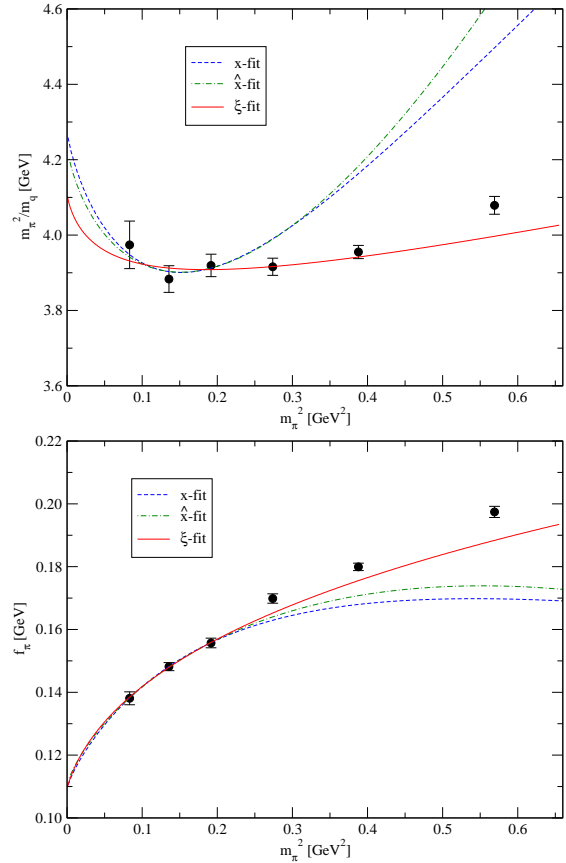

  \centering
  \includegraphics*[width=0.42\textwidth]{figure/mp2r_nlo.eps}
  \includegraphics*[width=0.42\textwidth]{figure/fps_nlo.eps}
  \caption{
    Comparison of chiral expansion in terms of $x$, $\hat{x}$ and
    $\xi$.
    The plots represent $m_\pi^2/m_q$ (left) and $f_\pi$ (right).
    Fits of the three lightest data points with the NLO ChPT formulae
    (\protect\ref{eq:chiral_exp_mpi}) and (\protect\ref{eq:chiral_exp_fpi})
    are shown. 
    Results are from \cite{Noaki:2008iy}.
  } 
  \label{fig:mpifpi}
\end{figure}

The $\chi$PT provides a systematic expansion in terms of small $m_\pi^2$
and $p^2$, but the region of convergence of this chiral expansion is not
known a priori. 
With the exact chiral symmetry, the test is conceptually cleanest, since no
additional terms to describe the violation of chiral symmetry has to be
introduced.
(With other fermion formulations, this is not the case as already noticed.
The unknown correction terms are often simply ignored.)

For the pion mass $m_\pi$ and decay constant $f_\pi$ the expansion is given as
\begin{eqnarray}
  \label{eq:chiral_exp_mpi}
  \frac{m_\pi^2}{m_q} & = & 2B
  \left[ 1 + \frac{1}{2} x\ln x + c_3 x + O(x^2) \right],
  \\
  \label{eq:chiral_exp_fpi}
  f_\pi & = & f
  \left[ 1 - x\ln x + c_4 x + O(x^2) \right],
\end{eqnarray}
where $m_\pi$ and $f_\pi$ denote the quantities after the corrections while 
$m$ and $f$ are those at the leading order.
The expansions (\ref{eq:chiral_exp_mpi}) and (\ref{eq:chiral_exp_fpi}) may be
written in terms of either
$x\equiv 2m^2/(4\pi f)^2$, 
$\hat{x}\equiv 2m_\pi^2/(4\pi f)^2$, or
$\xi \equiv 2m_\pi^2/(4\pi f_\pi)^2$
(we use a notation of $f_\pi$ = 131~MeV).
They all give an equivalent description at this order,
while the convergence behavior may depend on the expansion parameter.

Among other works, I use our own data (by the JLQCD and TWQCD
collaborations) for a discussion here.
Figure~\ref{fig:mpifpi} shows the comparison of different expansion
parameters in two-flavor QCD \cite{Noaki:2008iy}.
The fit curves are obtained by fitting three lightest data points with the
three expansion parameters, which provide equally precise
description of the data in the region of the fit. 
If we look at the heavier quark mass region, however, it is clear that only
the $\xi$-expansion gives a reasonable function and others miss the data
points largely.
This clearly demonstrates that at least for these quantities the convergence
of the chiral expansion is much better with the $\xi$-parameter
than with the other conventional choices.
This is understood as an effect of resummation of the chiral expansion
by the use of the ``renormalized'' quantities $m_\pi^2$ and $f_\pi$.
In fact, only with the $\xi$-expansion we could fit the data including
the kaon mass region with the next-to-next-to-leading order (NNLO) formulae
\cite{Noaki:2008iy}.

\begin{figure}[t]
  \centering
  \includegraphics*[width=0.49\textwidth]{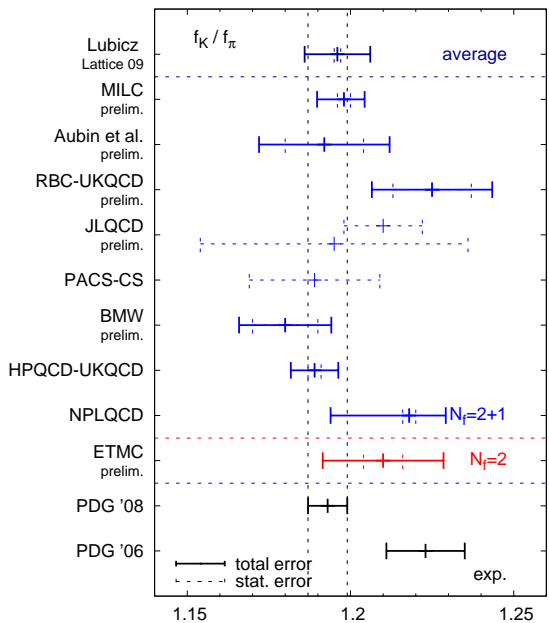}
  \caption{
    $f_K/f_\pi$ from various lattice groups.
    A plot from E.~Scholz \cite{Scholz:2009yz}.
    References for individual data can be found there.
    In the plot, PDG'08 is from the experimental data of leptonic kaon
    decay, assuming $|V_{us}|=0.2255(19)$.
  }
  \label{fig:fKfpi}
\end{figure}

With the NLO formulae (like those in
(\ref{eq:chiral_exp_mpi}) or (\ref{eq:chiral_exp_fpi})),
the convergence of the chiral expansion is marginal in the kaon mass
regime. 
In fact, some groups concluded not to use the SU(3) $\chi$PT for the
kaon sector but use the SU(2) formula with the strange quark treated
as a heavy particle.
This corresponds to an expansion in terms of $m_{ud}/m_s$ and is a
theoretically consistent treatment, though the predictive power of
$\chi$PT is lost to some extent.
A summary of the results for $f_K/f_\pi$ from a review talk at Lattice
2009 by Scholz \cite{Scholz:2009yz} is shown in Figure~\ref{fig:fKfpi}.
Different groups take different strategies on the treatment of strange
quark in the chiral fit, {\it i.e.} SU(2) or SU(3), NLO or NNLO.
At the level of the error of order $\pm$0.02, the results are in
agreement.

\subsection{Kaon semi-leptonic form factor}
Whether the strange quark can be treated within the SU(3) $\chi$PT is
an important issue, since the main sources of phenomenological
information for kaon physics relies on $\chi$PT.
For instance, the determination of a CKM matrix element $|V_{us}|$
uses the semileptonic decays $K\to\pi\ell\nu$.
It is made precise because the relevant form factor $f^+(q^2)$ is
normalized to 1 at $q^2=0$ in the degenerate SU(3) limit.
The deviation from there is estimated using the SU(3) $\chi$PT
\cite{Leutwyler:1984je}.  
Even the lattice calculations use its formula as a guide to fit the
data. 
The recent result by the RBC-UKQCD collaboration 
with 2+1 flavors of domain-wall fermions \cite{Boyle:2007qe} 
is $f^+(0)=0.964(5)$, which is consistent with 0.961(8) of
the early phenomenological work \cite{Leutwyler:1984je}.

\subsection{Neutral kaon mixing}
Results for $B_K$, the bag parameter to characterize the
$K^0-\bar{K}^0$ mixing, has been made very precise and stable by
recent lattice calculations. 
The recent results in 2+1-flavor QCD are:
$\hat{B}_K$ = 0.720(13)(37) with 2+1 flavors of domain-wall fermions
\cite{Antonio:2007pb}, 
0.724(8)(28) with domain-wall valence quarks over 2+1 flavors
of staggered sea quarks \cite{Aubin:2009jh}.
Even including two-flavor calculations using overlap
\cite{Aoki:2008ss} and twisted-mass \cite{Dimopoulos:2008hb} fermions,
all the dynamical lattice calculations give consistent results.
The average given by Lubicz at Lattice 2009 \cite{Lubicz_lat09} is 
$\hat{B}_K=0.731(7)(35)$.

With the 5\% error, $B_K$ is no longer the dominant uncertainty in
drawing the constraint on the CKM unitarity triangle from the indirect
CP violation parameter $|\epsilon_K|$.
Since $|\epsilon_K|$ is related to the $(\rho,\eta)$ parameters of the
CKM unitary triangle as
$|\epsilon_K|\propto B_KA^2\eta (c_1+c_2A^2(1-\rho))$ ($c_1$ and $c_2$
are known constants),
the uncertainty in $|V_{cb}|\equiv A\lambda^2$ is now more important, 
as emphasized in \cite{VandeWater:2009uc}.

For the lattice calculations related to the kaon physics, I also refer
the readers to a recent comprehensive review by Boyle at the Kaon09
conference \cite{Boyle:2009te}.

\vspace*{4mm}
\section{Fundamental parameters}
\label{sec:Fundamental_parameters}
Next, I discuss the recent determinations of fundamental parameters in
QCD, {\it i.e.} the strong coupling constant and quark masses.

\subsection{Strong coupling constant}

\begin{figure}[t]
  \centering
  \includegraphics[width=0.44\textwidth]{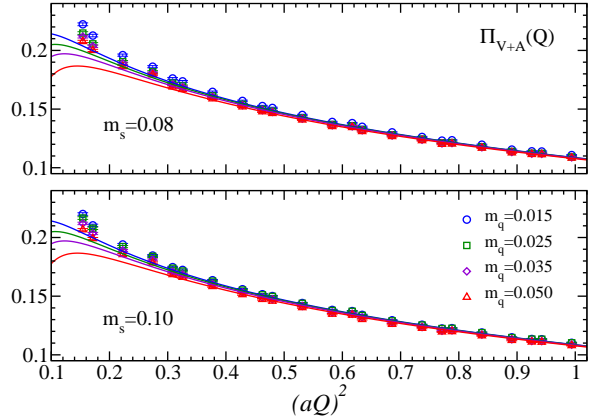}
  \caption{
    Fit of the vacuum polarization function on the lattice with
    continuum perturbative calculation.
    Data from the JLQCD collaboration with 2+1 flavors of dynamical
    overlap fermions \cite{Onogi_lat09}.
  }
  \label{fig:VPF}
\end{figure}

\begin{figure*}[tb!]
  \centering
  \includegraphics[width=0.80\textwidth]{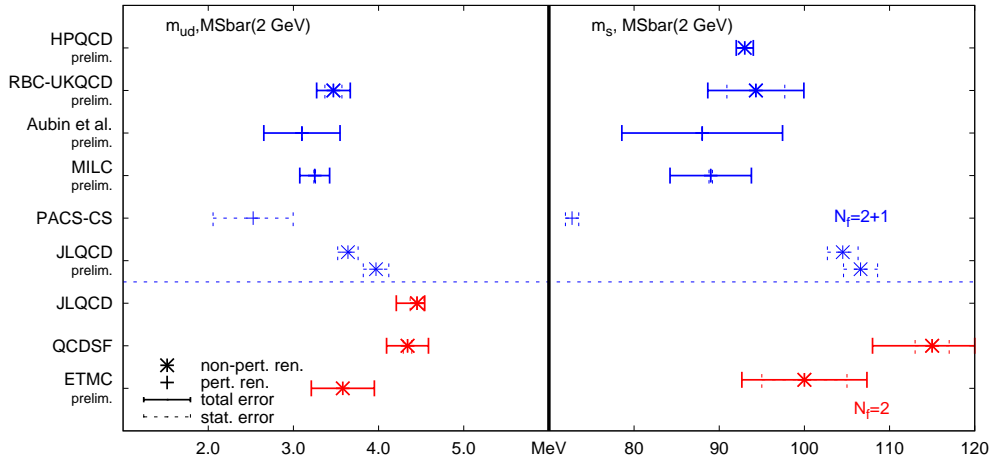}
  \caption{
    Summary of the quark mass determinations from the lattice.
    A plot from Scholz \cite{Scholz:2009yz}.
    References for individual data can be found there.
    Results with perturbative and non-perturbative renormalization
    factors are distinguished by pluses and starts, respectively.
  }
  \label{fig:quark_mass}
\end{figure*}

The determination of the strong coupling constant is equivalent to an
input of the lattice scale $1/a$.
This is because, for a given lattice with a bare coupling constant
$\alpha_s^{\mathrm{lat}}$, it gives the running 
of the coupling constant $\alpha_s^{\mathrm{lat}}(a^{-1})$.
But, in order to obtain the value in more familiar definitions,
such as the one in the $\overline{\mathrm{MS}}$ scheme,
one has to convert the result using perturbation theory as 
$\alpha_s^{\overline{\mathrm{MS}}}(\mu)=Z(\mu a)
 \alpha_s^{\mathrm{lat}}(a^{-1})$. 
Therefore, precise determination requires a good control of systematic
errors in the perturbative matching factor $Z(\mu a)$.
This is the reason that the previous lattice calculation by the HPQCD
collaboration \cite{Mason:2005zx,Davies:2008sw}
employed an automated perturbative calculation methods
to calculate two-loop contributions on the lattice.

Recently, new methods have been proposed.
They are based on a physical quantity which is sufficiently
short-distance and is calculable in both continuum and lattice
theories.
A well-known example is the Adler function, which is a derivative of
the vacuum polarization function
$D(Q^2)\equiv -Q^2 d\Pi_V(Q^2)/dQ^2$.
Since this quantity does not have an ultraviolet divergence, the
perturbative calculation in the continuum theory using the dimensional
regularization can be applied for the lattice data without
modifications except for possible discretization effects.
Therefore, one may fit the lattice data using the continuum formula
known to three-loops (or even four-loops for the leading term)
supplemented by an operator product expansion in $1/Q^2$.
Then, one can directly determine 
$\alpha_s^{\overline{\mathrm{MS}}}(\mu)$ \cite{Shintani:2008ga}.
The fit of lattice data in 2+1-flavor QCD is shown in
Figure~\ref{fig:VPF} for different light quark masses.
A preliminary result $\alpha_s(M_Z)=0.1181(8)(^{+4}_{-2})(^{+5}_{-6})$
has an error which is comparable to the previous best lattice
calculation, 0.1183(8) \cite{Mason:2005zx,Davies:2008sw}.

A closely related method is the use of the charmonium two-point
function.
By taking an appropriate moment in the coordinate space, the
ultraviolet divergence is removed and the lattice data can be directly
fitted with a corresponding continuum calculation.
The result is 
$\alpha_s(M_Z)=0.1174(12)$ \cite{Allison:2008xk}.
With this method, one can also determine the charm quark mass at the
same time \cite{Allison:2008xk}.

Overall, the lattice calculation now provides the most precise
determination of the strong coupling constant.
Along the lines of the recent calculations, further improvement is
expected.

\subsection{Quark masses}

Light quark masses can be extracted with pion and kaon masses as
inputs.
Therefore, the results depend on the fit functions (SU(2) or SU(3),
NLO or NNLO) and the mass range used in the analysis of pion and kaon
masses. 
Since the pole mass of quarks is not well defined due to confinement,
one has to use a renormalization scheme given at short distances such
as the $\overline{\mathrm{MS}}$ scheme.
Conversion from the lattice bare quark mass is done using perturbation
theory or partially using some non-perturbative methods.
The errors with purely perturbative method could be significantly
underestimated, as the coefficients of higher order terms are not known
a priori.
The results summarized in Figure~\ref{fig:quark_mass} (from the review
by Scholz \cite{Scholz:2009yz} at Lattice 2009) should be viewed with
these caveats in mind.
Since many results appeared only recently, the understanding of these
systematics will be achieved in the coming years.
I expect that the error is reduced to 2--5\% in the near future.

The isospin breaking, or the up and down quark mass ratio, cannot be
determined solely from the QCD calculation, as the electromagnetic
effect gives a substantial contribution.
For instance, the charged and neutral mass difference of pion is
dominated by the QED effect, which has to be subtracted to determine
$m_u-m_d$.
There is an attempt to identify the QED effect using the Weinberg-type
sum rule \cite{Das:1967it}:
$m_{\pi^\pm}^2-m_{\pi^0}^2 \propto \alpha_{EM}
\int_0^\infty dQ^2 Q^2 [\Pi_V(Q^2)-\Pi_A(Q^2)]$.
Here, the difference of the vacuum polarization functions in the
vector and axial-vector channels is involved, which means that this
mass difference is triggered by the spontaneous chiral symmetry
breaking.
With exact chiral symmetry, lattice calculation of
$\Pi_V(Q^2)-\Pi_A(Q^2)$ is possible \cite{Shintani:2008qe}.
Another interesting method to calculate the QED effect is a
simulation of the whole QCD+QED system \cite{Zhou:2008gb,Izubuchi_kaon09}.
Then one can also calculate the charged and neutral kaon mass
difference, which is useful to study the amount of the violation of the
Dashen's theorem
$m_{K^\pm}^2-m_{K^0}^2=m_{\pi^\pm}^2-m_{\pi^2}^2$
\cite{Dashen:1969eg}.
More striking application would be the mass difference between proton
and neutron, as it is related to a big question of why the matters in
Nature are stable.
For a recent analysis, see \cite{Izubuchi_kaon09}.

\vspace*{4mm}
\section{Heavy flavors}
\label{sec:Heavy_flavors}
The biggest motivation to consider the heavy quark flavors is to put
constraints on the CKM unitarity triangle.
Since the experiments of $B$ meson decays have been dramatically
improved over the last several years, the required precision for the
lattice calculation has become high.
Namely, to be interesting, the lattice calculation for the decay
constants, bag parameters, and form factors has to be typically as
precise as 5\% or even better, which is a challenge for the lattice
QCD community. 

\begin{figure}[t]
  \centering
  \includegraphics[width=0.48\textwidth]{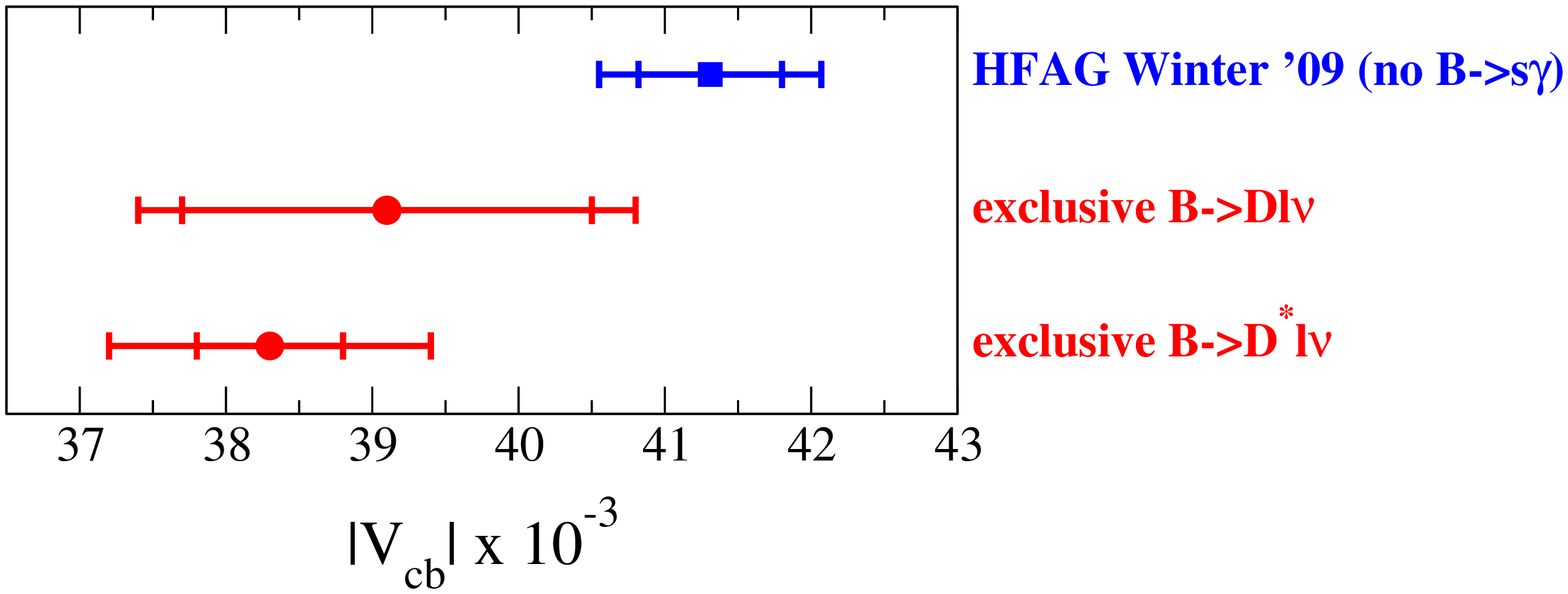}
  \\[2mm]
  \includegraphics[width=0.48\textwidth]{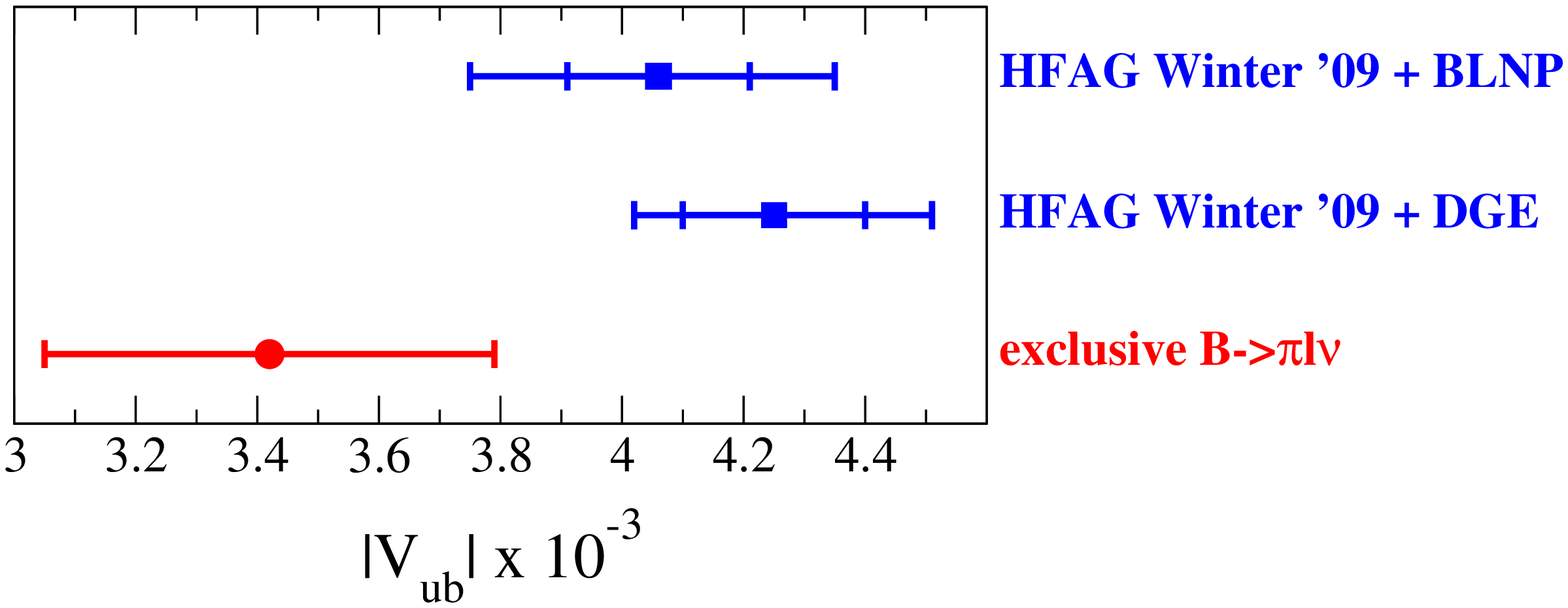}
  \caption{
    Inclusive and exclusive determinations of 
    $|V_{cb}|$ (top panel) and $|V_{ub}|$ (bottom panel).
    Plats are taken from Van de Water \cite{VandeWater:2009uc}.
    The lattice data are those in \cite{Bernard:2008dn}.
  }
  \label{fig:VcbVub}
\end{figure}

For instance, determination of the CKM matrix elements $|V_{cb}|$ and
$|V_{ub}|$ can be done through inclusive or exclusive decay modes.
The inclusive determination uses perturbative calculation relying on
the quark-hadron duality assumption, while the exclusive determination
requires the lattice calculation of the form factors.
So far, the precision of the inclusive determination is slightly
better for both $|V_{cb}|$ and $|V_{ub}|$, as far as the error is
taken as a face value.
See a summary plot in Figure~\ref{fig:VcbVub} made by Van de Water
\cite{VandeWater:2009uc}.
More importantly, the inclusive and exclusive determinations are
inconsistent with each other at around 2$\sigma$ level.
In order to understand this discrepancy, more precise calculations and
experiments are required.

Since the Compton wavelength of heavy quarks is too short to treat on
the lattice that is available with present computational power, one
has to use some effective theory.
The idea is to factor out a trivial heavy quark mass dependence
$e^{-im_Qt}$ from physical amplitudes; remaining terms are organized as
an expansion in terms of $1/m_Q$. 
The effective theories realize this in a language of effective
lagrangians, and it is also possible to construct a lattice version.
The problem is that the effective theory has many parameters that
have to be determined so that it reproduces the original QCD.
In many cases, the matching of these parameters is done by
perturbation theory, with which neglected higher order corrections
become a source of systematic error.
Non-perturbative parameter matching has also be attempted
\cite{Heitger:2003nj}, but is not common as it required dedicated
study for each choice of lattice gauge and fermion actions.
(For an introductory article on heavy quarks on the lattice, see
\cite{Hashimoto:2004fv}).

On the other hand, as the available computational resources increase,
it has also become possible to directly simulate heavy quarks with the 
lattice spacing kept as small as possible.
It is feasible only in the charm quark mass region, and an
extrapolation is necessary toward bottom quark mass.
In particular a variant of the staggered fermion (the HISQ action)
\cite{Follana:2006rc} 
has been found to be useful to minimize discretization effects.
Because the staggered fermion has a chiral symmetry (at a price of
broken flavor symmetry), the order of discretization effect is limited
to $O(a^2)$, $O(a^4)$, ...; the HISQ action is designed to remove the
leading $O(a^2)$ effect so that the remaining error is $O(a^4)$ at the
tree level.
Other non-chiral fermions have the errors of $O(a)$, $O(a^2)$, ...;
the removal of the lowest order effect leaves the $O(a^2)$
effect, which is still substantial.

\begin{figure}[t]
  \centering
  \includegraphics[width=0.46\textwidth]{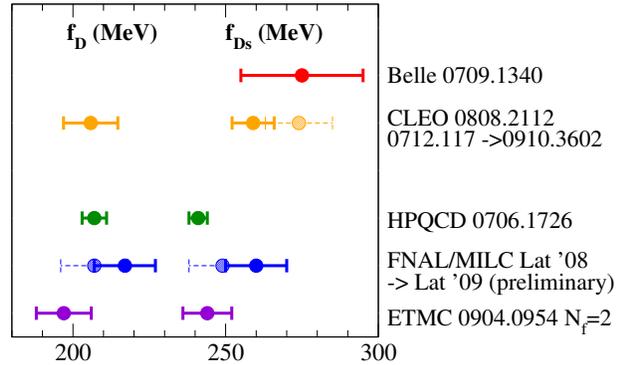}
  \caption{
    $D_{(s)}$ meson decay constant.
    Experimental data and lattice calculations are compared.
    A plot from Van~de~Water \cite{VandeWater:2009uc}.
    CLEO and FNAL/MILC data have been updated in 2009; the solid lines
    are the latest results.
  }
  \label{fig:fD}
\end{figure}

In order to demonstrate how precision is crucial in heavy flavor
phenomenology, I show a summary plot of the calculation of $D_{(s)}$
meson decay constant $f_{D_{(s)}}$ in Figure~\ref{fig:fD}.
Although there are three recent lattice calculations, the result of
the HPQCD collaboration \cite{Follana:2007uv} using the HISQ action 
has an order of magnitude smaller error quoted and thus dominates the 
lattice calculations.
If it is compared with experiments, one may conclude that there is
a significant discrepancy in $f_{D_s}$, which triggered theorists to
consider some new physics models that may explain this
\cite{Dobrescu:2008er,Kronfeld:2009cf}.
Therefore, it is very important to have other calculations that reach
this level of precision with different lattice fermion formulations.
This needs two or three more years in my opinion.

\vspace*{4mm}
\section{Summary and perspective}
The development of supercomputers was initiated in 1980s, and an
exponential growth of the computational power has been continued since
then. 
The rate is about a factor of 10 in 5 years over the last 25--30
years, and still no trend of speed-down is observed.
This, of course, has driven the rapid improvement of lattice
calculations. 
Until 1990s, only the quenched calculations were possible, and the
results were subject of uncontrolled systematic errors.
Large-scale dynamical fermion simulations were started at around 
late 90s, and with lots of efforts to improve algorithms and
techniques the simulations in a realistic setup have become feasible
recently. 
This means that the lattice calculation can now produce real
{\it predictions} from QCD for many interesting quantities.

Theoretically, the biggest achievement in the last 10--20 years in
lattice field theory is the formulation of lattice fermions with
chiral symmetry.
It removed a large class of limitations of lattice calculations.
Indeed, spontaneous chiral symmetry breaking and related phenomena can
now be directly simulated in lattice QCD.

These facts promise a wider range of applications of lattice QCD.
The coverage of this talk has been limited, but lattice QCD may be
useful in many other areas of particle and nuclear physics.
They include nucleon structure, spin physics, exotic hadrons, muon
$g-2$, flavor ($B$, $D$, and $K$) physics, heavy ion physics, and even
dark matter search or neutrino interactions
--- nearly all the subjects that were covered by the 
{\it Physics in Collision} conference.
Lattice QCD is your friend!

\section*{Acknowledgements}
I would like to thank Enno Scholz and Ruth Van de Water for allowing me
to use the plots they prepared for their Lattice 2009 plenary talks.
I also thank the members of JLQCD/TWQCD for fruitful collaborations. 
The author is supported in part by Grant-in-aid for Scientific
Research (No.~21674002).


\end{document}